\documentclass[journal=ancac3, manuscript=article]{achemso}
\setkeys{acs}{maxauthors=20, etalmode=truncate}
\usepackage[version=3]{mhchem} 
\usepackage[english]{babel}
\usepackage[utf8]{inputenc} 
\usepackage{siunitx} 
\usepackage{hyperref} 
\usepackage{xcolor} 
\usepackage{soul} 
\usepackage{changepage}
\usepackage{ulem}

\usepackage{amsmath}
\usepackage{graphicx}

\definecolor{delftdark}{rgb}{0, .5, .7}

\title{Tunable quantum dots from atomically precise graphene nanoribbons using a multi-gate architecture}

\author{\normalsize Jian Zhang}
\affiliation{\footnotesize Transport at Nanoscale Interfaces Laboratory, Empa, Swiss Federal Laboratories for Materials Science and Technology, 8600 Dübendorf, Switzerland}
\altaffiliation{\footnotesize These authors contributed equally to this work.}
\email{jian.zhang@empa.ch}

\author{Oliver Braun}
\affiliation{\footnotesize Transport at Nanoscale Interfaces Laboratory, Empa, Swiss Federal Laboratories for Materials Science and Technology, 8600 Dübendorf, Switzerland}
\alsoaffiliation{Department of Physics, University of Basel, 4056 Basel, Switzerland}
\altaffiliation{\footnotesize These authors contributed equally to this work.}

\author{Gabriela Borin Barin}
\affiliation{\footnotesize nanotech@surfaces Laboratory, Empa, Swiss Federal Laboratories for Materials Science and Technology, 8600 Dübendorf, Switzerland}

\author{Sara Sangtarash}
\affiliation{\footnotesize School of Engineering, University of Warwick,
Coventry CV4 7AL, United Kingdom}

\author{Jan Overbeck}
\affiliation{\footnotesize Transport at Nanoscale Interfaces Laboratory, Empa, Swiss Federal Laboratories for Materials Science and Technology, 8600 Dübendorf, Switzerland}
\alsoaffiliation{\footnotesize Department of Physics, University of Basel, 4056 Basel, Switzerland}

\author{Rimah Darawish}
\affiliation{\footnotesize nanotech@surfaces Laboratory, Empa, Swiss Federal Laboratories for Materials Science and Technology, 8600 Dübendorf, Switzerland}
\alsoaffiliation{\footnotesize Department of Chemistry and Biochemistry, University of Bern, 3012 Bern, Switzerland}

\author{Michael Stiefel}
\affiliation{\footnotesize Transport at Nanoscale Interfaces Laboratory, Empa, Swiss Federal Laboratories for Materials Science and Technology, 8600 Dübendorf, Switzerland}

\author{Roman Furrer}
\affiliation{\footnotesize Transport at Nanoscale Interfaces Laboratory, Empa, Swiss Federal Laboratories for Materials Science and Technology, 8600 Dübendorf, Switzerland}

\author{Antonis Olziersky}
\affiliation{\footnotesize IBM Research - Zurich, 8803 Rüschlikon, Switzerland}

\author{Klaus M\"ullen}
\affiliation{\footnotesize Max Planck Institute for Polymer Research, 55128 Mainz, Germany}

\author{Ivan Shorubalko}
\affiliation{\footnotesize Transport at Nanoscale Interfaces Laboratory, Empa, Swiss Federal Laboratories for Materials Science and Technology, 8600 Dübendorf, Switzerland}

\author{Abdalghani H.S. Daaoub}
\affiliation{\footnotesize School of Engineering, University of Warwick,
Coventry CV4 7AL, United Kingdom}

\author{Pascal Ruffieux}
\affiliation{\footnotesize nanotech@surfaces Laboratory, Empa, Swiss Federal Laboratories for Materials Science and Technology, 8600 Dübendorf, Switzerland}

\author{Roman Fasel}
\affiliation{\footnotesize nanotech@surfaces Laboratory, Empa, Swiss Federal Laboratories for Materials Science and Technology, 8600 Dübendorf, Switzerland}
\alsoaffiliation{\footnotesize Department of Chemistry, Biochemistry and Pharmaceutical Sciences, University of Bern, 3012 Bern, Switzerland}

\author{Hatef Sadeghi}
\affiliation{\footnotesize School of Engineering, University of Warwick,
Coventry CV4 7AL, United Kingdom}
\email{Hatef.Sadeghi@warwick.ac.uk}

\author{Mickael L. Perrin}
\affiliation{\footnotesize Transport at Nanoscale Interfaces Laboratory, Empa, Swiss Federal Laboratories for Materials Science and Technology, 8600 Dübendorf, Switzerland}
\alsoaffiliation{\footnotesize Department of Information Technology and Electrical Engineering, ETH Zurich, 8092 Zurich, Switzerland}
\email{mickael.perrin@empa.ch}

\author{Michel Calame}
\affiliation{\footnotesize Transport at Nanoscale Interfaces Laboratory, Empa, Swiss Federal Laboratories for Materials Science and Technology, 8600 Dübendorf, Switzerland}
\alsoaffiliation{\footnotesize Department of Physics, University of Basel, 4056 Basel, Switzerland}
\alsoaffiliation{\footnotesize Swiss Nanoscience Institute, University of Basel, 4056 Basel, Switzerland}
\email{michel.calame@empa.ch}

\begin{document}



\section*{Abstract}
\textbf{Atomically precise graphene nanoribbons (GNRs) are increasingly attracting interest due to their largely modifiable electronic properties, which can be tailored by controlling their width and edge structure during chemical synthesis. In recent years, the exploitation of GNR properties for electronic devices has focused on GNR integration into field-effect-transistor (FET) geometries. However, such FET devices have limited electrostatic tunability due to the presence of a single gate. Here, we report on the device integration of 9-atom wide armchair graphene nanoribbons (9-AGNRs) into a multi-gate FET geometry, consisting of an ultra-narrow finger gate and two side gates. We use high-resolution electron-beam lithography (EBL) for defining finger gates as narrow as 12~nm and combine them with graphene electrodes for contacting the GNRs. Low-temperature transport spectroscopy measurements reveal quantum dot (QD) behavior with rich Coulomb diamond patterns, suggesting that the GNRs form QDs that are connected both in series and in parallel. Moreover, we show that the additional gates enable differential tuning of the QDs in the nanojunction, providing the first step towards multi-gate control of GNR-based multi-dot systems.}\\


Bottom-up synthesized GNRs have attracted considerable interest as possible future electronic building blocks. This is mainly due to the fact their chemical structure can be controlled with atomic precision, a property that top-down etched GNRs lack.\citep{gunlycke2007room, stampfer2009energy,Xiujun2022Minimizing} Bottom-up synthesized GNRs can, therefore, be regarded as a \textit{designer quantum material}, where the material properties can be designed by selecting the appropriate chemical precursors and synthetic routes\cite{Cai2010AtomicallyPreciseBottom, Liu2015CoveEdgedLow, Cai2014GrapheneNanoribbonHeterojunctions,Ruffieux2016surfaceSynthesisGraphene,Groning2018EngineeringRobustTopological,Rizzo2018TopologicalBandEngineering,Sun2020MassiveDiracFermion,Yamaguchi2020Smallbandgapatomically,Chen2020GrapheneNanoribbonssurface,Li2020FjordEdgeGraphene,Pawlak2020BottomSynthesisNitrogen,Sun2020CoupledSpinStates,Cirera2020TailoringTopologicalOrder, Yang2022Solution}. As such, one can largely tune their bandgap, \citep{Son2006EnergyGapsGraphene,Kimouche2015Ultranarrowmetallic} form $pn$-junctions within a single, heterogeneous ribbon,\citep{Cai2014GrapheneNanoribbonHeterojunctions} tailor spin-polarized states\cite{wang2016giant,Ruffieux2016surfaceSynthesisGraphene} and even topologically non-trivial phases.\citep{Cao2017TopologicalPhasesGraphene, Rizzo2018TopologicalBandEngineering, Groning2018EngineeringRobustTopological, Sun2020MassiveDiracFermion,Cirera2020TailoringTopologicalOrder} 
Exploiting these properties in electronic devices requires contacting strategies that preserve the integrity of the GNRs, while at the same time allowing for charge carriers to flow through. Moreover, many technological applications require electrostatic control over the level structure of the GNRs. For example, field-effect transistors require the presence of a single gate electrode to tune the channel to conductance, while multiple gate electrodes are needed for the realization of qubits.  

Several prototypical GNR devices have been studied to date \cite{Llinas2017ShortchannelField,Martini2019StructuredependentElectrical,Passi2018FieldeffectTransistors,ElAbbassi2020ControlledQuantumDot,Sun2020MassiveDiracFermion,Richter2020ChargeTransportMechanism,braun2021optimized,senkovskiy2021tunneling}, exhibiting various charge-transport characteristics, such as high-performance field-effect transistors operating at room temperature\cite{Llinas2017ShortchannelField}, gate-tunable QDs at low temperature\cite{ElAbbassi2020ControlledQuantumDot,Sun2020MassiveDiracFermion} and temperature-activated transport through micron-sized films\cite{Richter2020ChargeTransportMechanism, senkovskiy2021tunneling}. 
However, many challenges remain in the device integration of these materials. On the one hand, the contacts need to be improved further\cite{saraswat2021materials}, as well as the transfer process from the growth substrate to the devices substrate which can lead to defects, impurities, and adsorbates at the interface between GNRs and the electrode material. On the other hand, advanced gating strategies, such as ultrashort transistors\citep{wu2022vertical,Jiang2020Ultrashort} or multi-gate architectures, are highly desirable for devices that require additional control over the electrostatic landscape of the device. To date, due the nanoscale size of the GNRs, only field-effect-transistor devices have been realized\cite{Llinas2017ShortchannelField,Martini2019StructuredependentElectrical,Passi2018FieldeffectTransistors,ElAbbassi2020ControlledQuantumDot,Sun2020MassiveDiracFermion,Richter2020ChargeTransportMechanism,braun2021optimized,senkovskiy2021tunneling}. More advanced device architectures with multiple gates that are individually addressed require a very high control over the fabrication of the multiple gates, the electrodes, and the alignment between them.

Here, we report on the integration of GNRs into a multi-gate field-effect transistor with graphene electrodes. Our device design consists of a narrow finger gate and two additional side gates. This geometry improves gating capabilities by allowing for the generation of an asymmetric gate field using the side gates. As such, the different sides of the nanogaps experience a different gate field, providing additional control over the electrostatic landscape of the junction.
The narrow gate is $\sim$10~nm in length, with an effective channel length of $<$15~nm, and is fabricated using CMOS-compatible processing steps. The graphene electrodes are created using EBL, which has a major advantage of the control of the nanogap position\cite{braun2021optimized} and a proper alignment with the underlying gates. This is in contrast to electrodes created using the electric breakdown procedure that has been commonly used for graphene.\citep{ElAbbassi2019RobustGraphenebased,ElAbbassi2020ControlledQuantumDot,Sun2020MassiveDiracFermion} Moreover, our fabrication protocol allows for the integration of the GNRs at the very last stage of device fabrication. Similar approaches with the integration of the sensitive material in the final step have been shown to lead to major improvements in the device performance, as demonstrated for example for MoS$_{2}$.\citep{radisavljevic2011single,marega2020logic,liu2021transferred} The design of the devices is supported by finite-element calculations for optimizing the various geometrical parameters and maximizing the effective electrostatic potential at the GNRs. Furthermore, low-temperature transport spectroscopy measurements reveal quantum dot (QD) behavior with addition energies in the range of 20-150~meV, and transport characteristics that are tunable using the two side gates. Our observations are supported by simple model calculations that highlight the importance of the asymmetric gate field in the junction area.

\section{Results}
\subsection{Devices design and fabrication}
\begin{figure}
\begin{center}
	  \includegraphics[width=0.99\textwidth]{ 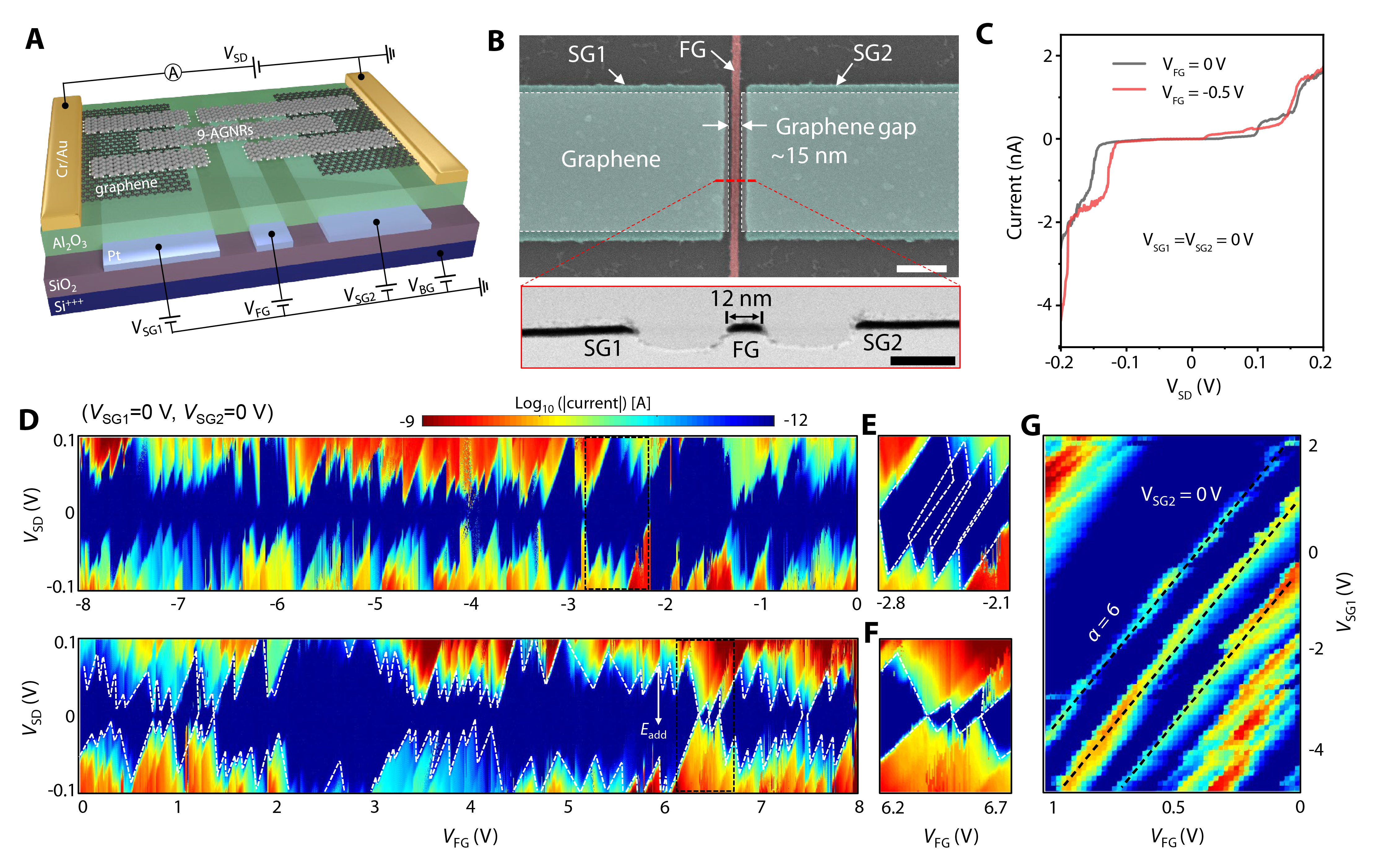} 
\end{center}
\caption{\textbf{Multi-gate 9-AGNRs quantum dot device.}
    (\textbf{a}) Schematic illustration of the device architecture, including the measurement circuit. 
    (\textbf{b}) False colored scanning electron micrograph image of a device prior to GNR transfer showing FG (red), SG1, and SG2 (green) below the graphene (white dashed line).Scale bar, 100~nm.  The inset presents a high-resolution transmission electron micrograph of a device cross section revealing the FG length and the separation to the side gates. Scale bar, 20~nm
    (\textbf{c}) Selected I-V characteristics at two different finger gate voltages recorded at T = 9~K.
  (\textbf{d}) Device 1: Recorded map of the current as a function of bias voltage (V$_{\textnormal{SD}}$) and finger gate voltage (V$_{\textnormal{FG}}$) obtained at T = 9~K. Coulomb diamonds are visible over a gate range of 16~V. High-resolution Coulomb diamonds, where dotted white lines indicate the diamond edges. 
    (\textbf{e}) Non-closing Coulomb diamonds and (\textbf{f}) closing Coulomb diamonds.  
    (\textbf{g}) Measured current at a fixed bias voltage of V$_{\textnormal{SD}}$= 0.1~V as a function of applied finger gate and side gate voltage. Dashed black lines are guides to the eye.
    }
    \label{fig:figure1}
\end{figure}

A schematic of the proposed device architecture is shown in Figure~\ref{fig:figure1}a. The finger-gate (FG) with nanometer-scale dimensions is fine-patterned under the 9-AGNRs junction, while two side-gates (SG1 and SG2) are defined under the source and drain graphene electrodes, respectively. As the electronic coupling between the GNRs and the graphene is weak, we anticipate the formation of quantum dots (QD) at low temperatures.\cite{ElAbbassi2020ControlledQuantumDot,Sun2020MassiveDiracFermion} The side gates (SG1 or SG2), as they are located close to the nanojunction, are employed to introduce an asymmetric electric gate field that will couple to the QD. We note that the interplay between the gate length, gate separation, gate oxide thickness and the applied potentials is very delicate to optimize to achieve a homogeneous electrostatic potential over the complete channel length. Thinner oxides lead to higher gate coupling but also lower breakdown voltages between the various gates. Moreover, reducing the distance between the gates also increases the screening of the gate potential by the neighboring gates. To investigate this balance, we performed finite-element calculations using Comsol Multiphysics. In Section 1 of the Supporting Information, we present the effective potential at the GNRs for various thicknesses of the Al$_{2}$O$_{3}$ and the gate separation. Graphene is modeled as a surface charge density; its value is calculated using the voltage applied on the gate located below the respective electrode, and the sum of quantum capacitance and geometric capacitance.

We show that thinner oxides down to 12~nm are generally more beneficial. A further result is that reducing the FG-SG separation is beneficial, but only down to 10~nm. Beyond that point, the field exceeds 1~V/nm, a strength where a breakdown of the oxide is likely to happen.\citep{wu2007current}

The starting point of the sample fabrication is a highly-doped silicon (Si) carrier chip with a 285~nm thick silicon dioxide (SiO$_{2}$), such that the Si substrate acts as a global back gate (BG). The finger gates are patterned on top as follows. First an 8~nm platinum (Pt) film is deposited using electron-beam evaporation. Then, a negative resist hydrogen silesquioxane (HSQ) is spin-coated as a resist to define the etch mask. This resist turns into SiO$_{2}$-like after EBL exposure and development, leading to a highly-resistive etch mask. A subsequent Ar$^{+}$-ion milling step transfers the etch mask feature to the metal film, separating the finger from the side gates. This process leads to very sharp features as it is not limited by grain sizes or other edges effects which are common when using electron-beam evaporation with a lift-off process. Our approach results in a finger gate with a length of 10-15~nm and a width of 500~nm. The nanometer-scale dimension of the FG ($<d$) allows for creating an ultra-short effective channel length while minimizing parasitic gate to source-drain capacitance. The 9-AGNRs junction is electrically isolated from the metal gates using a 30~nm thick aluminum oxide (Al$_{2}$O$_{3}$). The graphene electrodes are separated by a nanogap formed with high-resolution patterning by using EBL, as reported elsewhere.\citep{braun2021optimized} Here, the electrode separation $d$ is set to be $\sim$15~nm, large enough to eliminate direct tunneling contributions between the electrodes, but smaller than the average length of the 9-AGNRs. \citep{ DiGiovannantonio2018surfaceGrowthDynamics} In Section 2 of the Supporting Information, a more detailed description of the fabrication process is given.A scanning electron micrograph (SEM) of the final device before GNR transfer is presented in Fig.~\ref{fig:figure1}b, alongside a transmission electron micrograph (TEM). The image shows that the FG length is $\sim$12~nm, and the separation between the gates $\sim$30~nm. 

\subsection{Quantum dot formation at low temperature}
\label{sec:QD_LT}
Prior to the deposition of the GNRs, the nanogaps were electrically characterized to ensure a clear separation between the electrodes. Devices with currents $>$10~pA at V$_{\textnormal{SD}}$ = 4~V were excluded from further characterization (See Section 3.1 of the Supporting Information). Uniaxially aligned medium-density 9-AGNRs were synthesized on an Au(788) single crystal under ultrahigh vacuum conditions (See Materials and Methods).\citep{Overbeck2019OptimizedSubstratesMeasurement} A representative scanning tunneling micrograph is presented in Section 4 of the Supporting Information. The average GNR density is about 2 GNRs per terrace (3-4 nm in width), with an average length between 40-45~nm.\citep{DiGiovannantonio2018surfaceGrowthDynamics} The 9-AGNRs were then transferred to the multi-gate device substrate with the predefined and characterized graphene electrodes using a PMMA-based electrochemical delamination process.\citep{Overbeck2019OptimizedSubstratesMeasurement, senkovskiy2017making, Overbeck2019UniversalLengthdependent} In order to improve the device performance, a thermal annealing step was performed.\citep{braun2021optimized} After this thermal annealing step the integrity of the 9-AGNRs and their alignment with respect to the source-drain axis was confirmed using polarization-dependent Raman spectroscopy (See Section 4 of the Supporting Information). 

To evaluate the electrical properties of the 9-AGNRs after device integration, we recorded stability diagrams (current-voltage characteristics (IV) for varying FG voltage V$_{\textnormal{FG}}$ at T = 9~K). Two IVs are presented in Figure~\ref{fig:figure1}c for different finger gate voltages, both exhibiting blocked current at low bias voltage and multiple steps in current for increase bias voltages. These steps are typical for quantum dot behavior\cite{Kouwenhoven1997}. 
Figure~\ref{fig:figure1}d shows a stability diagram recorded with the voltages on SG1, SG2 and BG all set to 0~V. The plot shows irregular and aperiodic Coulomb diamonds (CD) over a FG voltage range of 16~V (-8~V to 8~V), with addition energies ($E_{\textnormal{add}}$) varying between 20~meV and 150~meV. Given the nanometer size of the GNRs, and in particular their extremely narrow width of $\approx$ 1~nm, we attribute these Coulomb diamonds to the formation of quantum dots in the GNRs with a discrete level structure\cite{Son2006EnergyGapsGraphene,MerinoDiez2017WidthdependentBand}.
The lever arm for the FG, $\alpha_{FG}$ = $\Delta V_{SD}$/$\Delta V_{FG}$, is determined to be in the range of 220-340 mV/V from the Coulomb diamonds, indicating a very strong gate coupling of the finger gate to the 9-AGNRs. For a large portion of the FG range, no crossing of the corresponding energy level with the Fermi energy of the electrodes is observed, i.e., no resonance is visible at zero bias. This behavior is highlighted in Fig.~\ref{fig:figure1}e for a gate range from -2.8~V to -2.1~V and is attributed to transport through two weakly-coupled QDs in series. Here, the serial QDs correspond most likely to different GNRs, but possibly also to the formation of localized states within a single GNR due to the presence of bite defects.\citep{Pizzochero2021Quantumelectronictransport} In other gate regimes, for example in Fig.~\ref{fig:figure1}f, the Coulomb diamonds are closing, indicating that in this gate range a single level of a QD is dominating the transport. Although not very likely, this feature could also be associated with multiple quantum dots in series with very similar energies. We also observe multiple overlapping diamonds with different addition energies, suggesting that charge transport occurs through two or more parallel QDs, presumably individual GNRs, with different energy spectra.\citep{ionicainfluence} These differences could be caused by different GNR length and local environments. The stability diagrams of devices 2-4 show similar features as Device 1 (see Supporting Information Section 3.2).\\

To investigate the additional control offered by the side gates, we recorded a map of the current as a function of V$_{\textnormal{FG}}$ and V$_{\textnormal{SG1}}$ for a fixed source-drain voltage V$_{\textnormal{SD}}$ = 100~mV, as shown in Fig.~\ref{fig:figure1}g. The plot displays multiple regions of high current that shift with the gate voltages and appear as lines. Three such lines of enhanced current are marked with black dashed lines. From the slopes we extract the relative gate coupling $\alpha$ between FG and SG1 and find $\alpha = \alpha_{\textnormal{FG}} /\alpha_{\textnormal{SG}} \approx$5.98, revealing that the FG couples more strongly to the GNR than the SGs. In this gate range, we do not observe any distinct signature of a double quantum dot system, with access to multiple charge states. We attribute this to the presence of multiple GNRs in the channel. In section 1 of the Supporting Information, we perform finite-element calculations to model the electrostatic potential in the junction area as a function of the voltages applied on the side gates and finger gate. Here, a relative gate coupling $\alpha$ of around $~\sim$10 is found, close to the experimentally observed value of $\sim$6. The difference between experiment and theory may be caused by local variations in the sample geometry, such as the gate and/or the gate oxide, or a slight misalignment of the graphene gap with respect to the finger gate.

\begin{figure}
\begin{center}\includegraphics[width=\textwidth]{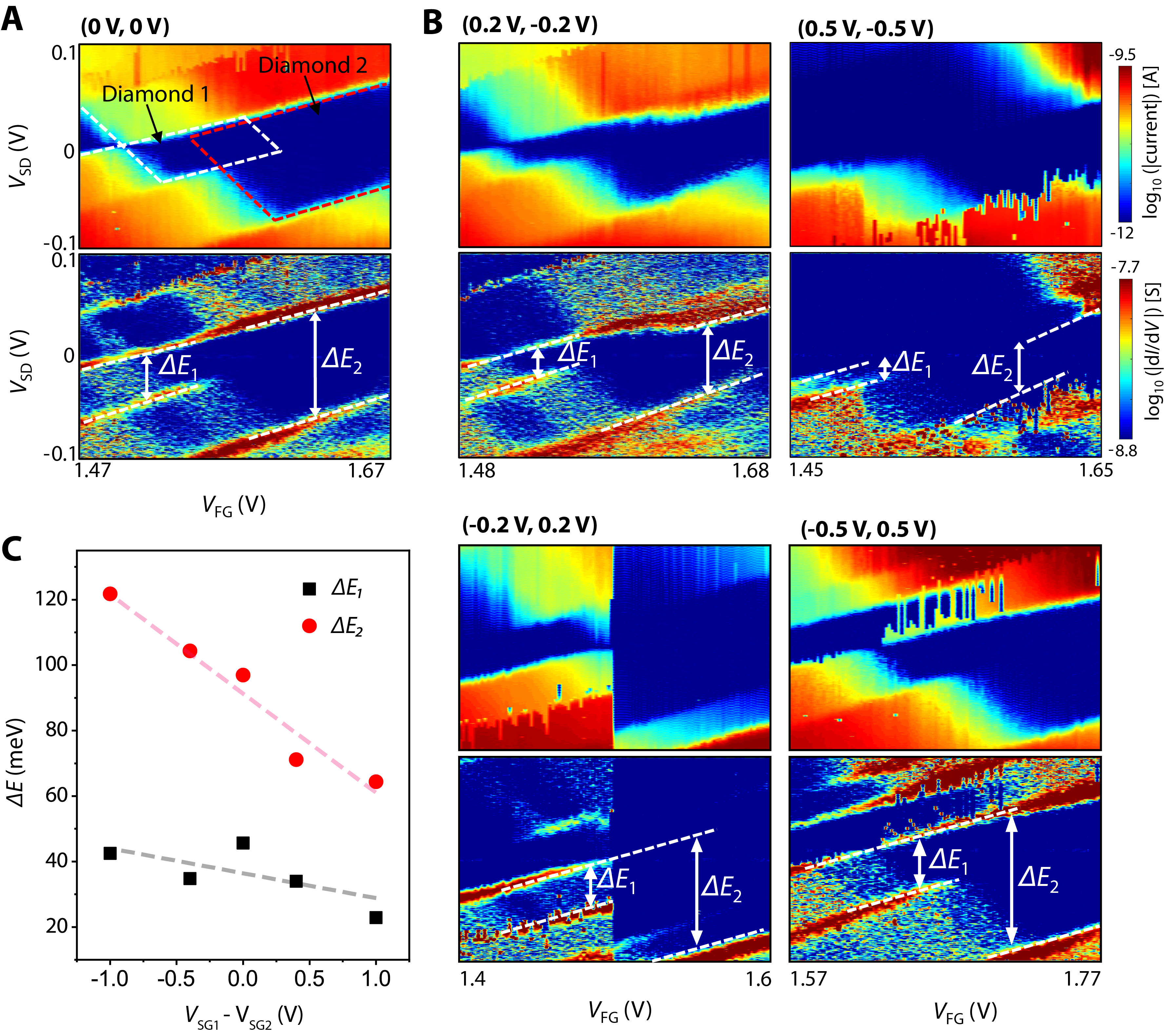}
          \end{center}
\caption{\textbf{Side-gate enabled tunability of the QDs.} 
(\textbf{a}) Top panel: Current map as a function of bias and finger gate voltage measured at T = 9~K, for 0~V applied on the side gates. As a guide to the eye, two diamond-shape areas are highlighted with dashed white and red lines, respectively. Bottom panels: Corresponding dI/dV maps. The energy difference $\Delta$E1 and $\Delta$E2 are highlighted by the white arrows.
(\textbf{b}) Same plots as in (A), but for different side gate voltages, as indicated in brackets on the top as (V$_{\textnormal{SG1}}$, V$_{\textnormal{SG2}}$). 
 (\textbf{c}) Extracted energies difference $\Delta$E1 and $\Delta$E2 for various side-gate voltage differences (V$_{\textnormal{SG1}}$ - V$_{\textnormal{SG2}}$). The red and black dash lines are linear fits to the data.
  }
    \label{fig:figure2}
\end{figure}

\subsection{Effect of side gates}
\label{subsec:QD_gating_efficiency}
In Fig.~\ref{fig:figure2}, we further characterize the effect of the side gates, and in particular on the electronic structure of the QDs formed in the device. We measure the current maps as a function of the finger gate voltage V$_{\textnormal{FG}}$ for different side gate voltages V$_{\textnormal{SG1}}$ and V$_{\textnormal{SG2}}$ (later indicated in brackets). The top panel of Fig.~\ref{fig:figure2}a presents the (0~V, 0~V) case, focusing on the finger gate voltage range around +1.6~V, where two diamonds of different sizes overlap, indicating two QDs with different physical size are weakly coupled and contacted in series. The lower panel shows the dI/dV plots, from which energy differences between the energy levels for different dots are identified, as indicated by $\Delta E_{1}$ and $\Delta E_{2}$. We note that these energies do not correspond to the addition energies of the two QDs. To extract the relative gating of the two QDs by the side gates, we perform the same stability diagram but now for different values of the side gates. To maximize the asymmetry of the electric field introduced in the nanojunctions, we apply voltages of opposite polarities on the two side gates. In Fig.~\ref{fig:figure2}b, we plot the current and dI/dV maps for four different side gate combinations. The figures show that the side gates significantly alter the shape and size of the two diamonds. Fig.~\ref{fig:figure2}c shows the energy difference $\Delta E_{1}$ and $\Delta E_{2}$ as a function of the voltage difference between V$_{\textnormal{SG1}}$ and V$_{\textnormal{SG2}}$. The plots show that $\Delta E_{1}$ is strongly modulated by the asymmetric gate field, with a shift of about 30~meV / 1~V. $\Delta E_{2}$, on the other hand, has a shift of about 10~meV / 1~V. This observation suggests that the asymmetric field introduced by the side gates leads to differential gating of one QD versus the other. This also implied that one of the QD is closer to one of the side gates.

\begin{figure}
    \begin{center}\begin{tabular}{l}
	  \includegraphics[width=0.99\textwidth]{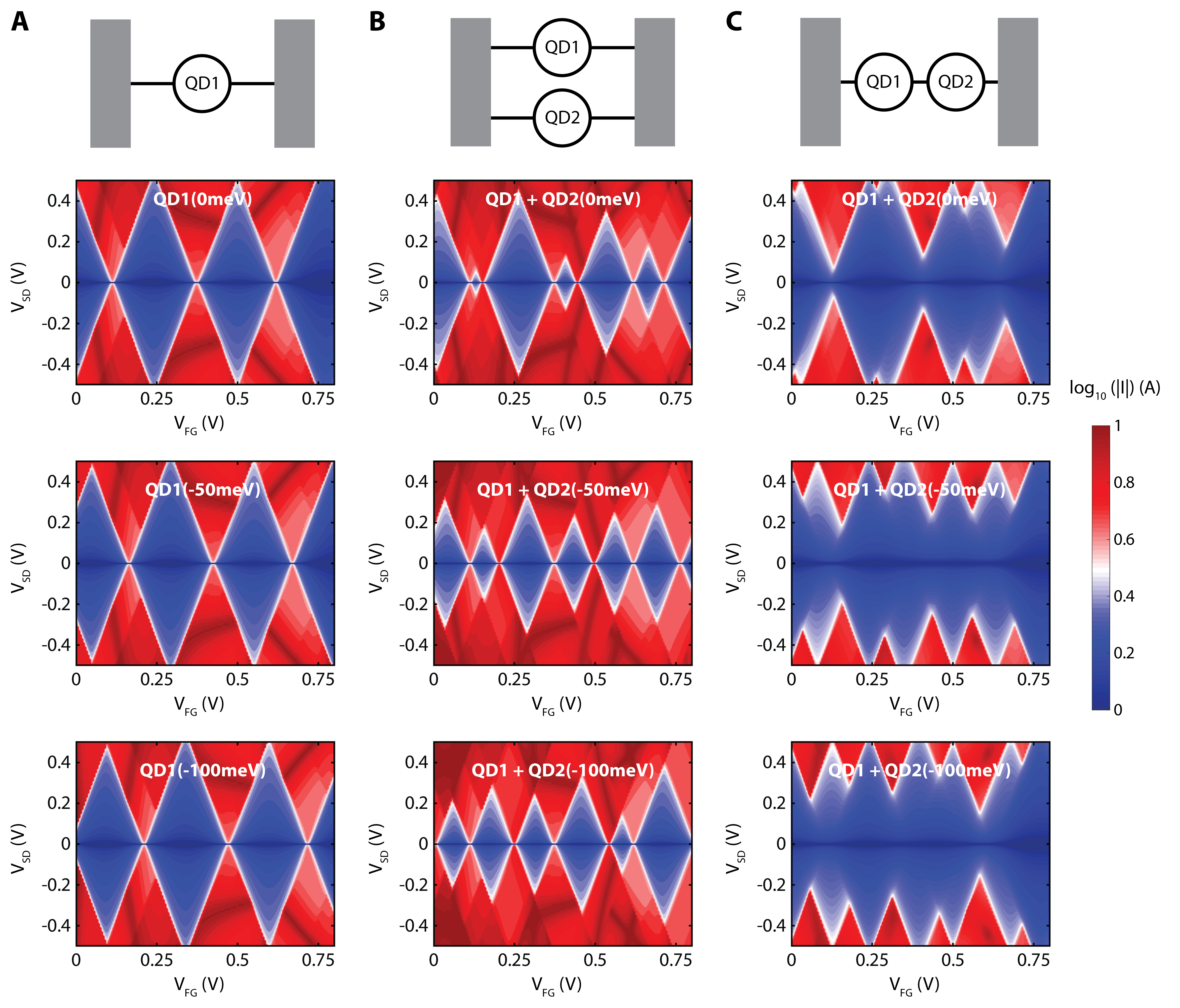} 
           \end{tabular}\end{center}
    \caption{\textbf{Model calculations.}
Computed current maps for a single QD (\textbf{a}), two QDs in parallel (\textbf{b}) and two QDs in series (\textbf{c}). The respective top stability diagrams show the current without any side gate voltages. To illustrate the role of the asymmetry introduced by the side gates, the middle and lower stability diagrams show the computed current maps for QD2 (for \textbf{b} and \textbf{c})  shifted by 50~meV and 100~meV, respectively.
    }
    \label{fig:figure3}
\end{figure}

To rationalize the experimental observations shown in Fig.~\ref{fig:figure1} and Fig.~\ref{fig:figure2}, we performed quantum transport calculations\cite{Sadeghi2018TheoryElectronPhonon} through single and multiple dots in series and parallel using a simple tight-binding model (see details in the Materials and Methods, and in supporting information Section 5). The two QDs have each a different level structure with addition energies that are nonetheless comparable in size. We assume that the main bias potential drop happens at the contact point to electrodes and remains constant over the QDs. This is a good approximation because the coupling to electrodes is weak. In Fig.~\ref{fig:figure3}a, we show the single QD case, with three current vs. FG voltage maps (stability diagrams). The top current map shows the current modulation by only the FG voltage. In the middle and bottom current map, to mimic the electrostatic potential change induced by the two side gates, the overall energy spectrum of the QD is shifted by -50 and -100~meV, respectively. We note that the effective potential experienced by the QD as a result of the two side gates will depend on its position in the nanojunction and the applied side gate voltages. Figure~\ref{fig:figure3}b and Fig.~\ref{fig:figure3}c present the same plots, but for two QDs in parallel and in series, respectively. Here, the asymmetry is modelled by moving the energy levels of QD2 with to those of QD1. As the two QDs have a comparable level structure, the stability diagrams for transport in parallel and series appear to be fairly similar. Nevertheless, current in parallel is about twice as large and an additional crossing point appears very close to those already present for the single QD case. For the series case, the current is twice as small and some crossing points are fading out. The situation becomes very different when QD2 is being shifted. In the parallel case, additional diamonds appear and grow in size, while the existing ones are being reduced. In series, the crossing points fade out even more until the diamonds do not close anymore. These plots indicate that the energy shift of one of the QDs in series with respect to the other, as induced by an asymmetric gate field, can drastically alter the electronic properties of the nano-junctions.

\section{Conclusion and Discussion}
\label{sec:local_gate_outlook} 

We developed a multi-gate FET geometry to contact uniaxially aligned 9-AGNRs that are contacted by graphene electrodes. At low temperatures, Coulomb blockade and single-electron tunneling is observed, with a very strong gate coupling of the finger gate to the 9-AGNRs, up to 340~meV/V. In addition, we demonstrate the additional tunability offered by the two side gates that are present in our device architecture and allow for differential electrostatic tuning of the multiple QDs present in the junction. 

The observed addition energies of $\sim$0.065~eV are in discrepancy with the DFT-calculated bandgap of 0.73~eV \citep{Son2006EnergyGapsGraphene} as well as the measured bandgap using scanning tunneling spectroscopy (STS) of 1.40~eV\citep{Talirz2017surfaceSynthesisCharacterization}. However, in our measurements, many QDs are presumably bridging the two electrodes in parallel and series, making the task of extracting the exact addition energy challenging. In particular, as observed in Fig.~\ref{fig:figure3}b, the effective diamond sizes is reduced when multiple QDs are connected in parallel, leading to an underestimation of the addition energies of the individual QDs. The multiple GNRs in the nanojunction also make it such that the QDs in series do not form an ideal double-dot system, in which each gate tunes primarily the levels of the closest QD. This aspect is further highlighted by the fact that different quantum dots couple differently to the different gates, illustrating the challenges in the exact positioning of the nanoscale gates, nanogaps, and GNRs with respect to each other.

Further efforts in reducing the electrode size for contacting fewer GNRs would be valuable for characterizing the physical properties of a single GNR integrated into devices, possibly by downscaling the width of the graphene electrode, or by using electrode materials that are naturally close to the atomic level, for example, carbon nanotubes\cite{Zhang2022Contactingatomicallyprecise}. Along similar lines, our approach might be further pushed towards an ultimately-narrow gate by using a carbon nanotube\citep{desai2016mos2} or even the edge of a graphene sheet\cite{wu2022vertical} as gate electrodes, as demonstrated for MoS$_{2}$-FETs. Finally, several charge jumps are observed in our measured stability diagrams, highlighting the fact that the effect of the substrate needs to be better controlled, for example, via encapsulation in hBN.

Overall, the differential gating of the QDs using multiple gates is a major step forward in the exploration of the transport characteristics of GNRs and their exploitation in future electronic device architectures.\cite{wang2021graphene} With further development, more advanced electronic quantum devices may be envisioned in which several gates are required to achieve the required control over the electrostatic landscape. In particular, the controlled formation of double quantum dot systems based on GNRs offers exciting prospects for the realization of GNR-based quantum technologies such as spin qubits that possess operation conditions above dilution refrigerators temperatures.

\section*{Materials and Methods}
\subsection*{Fabrication of multi-gate device substrates}
A detailed description is given in Supporting Information Section 2. In brief, a highly $p$-doped silicon (500~$\mu$m) with a thermally grown silicon dioxide layer (285~nm) is used as a base substrate. A titanium/platinum thin film (1~nm/8~nm) is evaporated on top and a negative resist hydrogen silesquioxane (HSQ) is spin-coated as a resist to define the etch mask. This resist turns into SiO$_{2}$-like film after EBL exposure and development, leading to a highly-resistive etch mask. A subsequent Ar$^{+}$-ion milling step transfers the etch mask feature to the metal film, separating the finger from the side gates.

\subsection*{Graphene growth, transfer, and nanogap formation}
Polycrystalline graphene is synthesized via chemical vapor deposition (CVD), transferred, and pre-patterned as reported elsewhere.\citep{braun2021optimized} After a first prepatterning of the graphene, a 60~nm thick CSAR resist (AR-P 6200.04, Allresist GmbH) is spin-coated. Following the second electron beam exposure, the resist is developed using a suitable developer (AR 600-546, Allresist GmbH) at room temperature for 1~min followed by an IPA rinse. Reactive ion etching, RIE (15~sccm Ar, 30~sccm O$_{\textnormal{2}}$, 25~W, 18~mTorr) for 6-8~s was used to cut the graphene within the CSAR gap. After RIE, the etching mask is removed by immersing in 1-Methyl-2-pyrrolidinone (NMP) (Sigma Aldrich) at room temperature for 10~min followed by 60~min at 80~$^\circ$C, cooled down for 30~min, rinsed with IPA, and blown dry with N$_{2}$. This approach yield clean and well-separated graphene electrodes ($<$15~nm nanogaps). 

\subsection*{On-surface synthesis of aligned 9-AGNRs and transfer to device substrate}
9-AGNRs were synthesized from 3',6'-diiodo-1,1':2',1"-terphenyl (DITP). \citep{DiGiovannantonio2018surfaceGrowthDynamics} Using a Au(788) single crystal (MaTeK, Germany) as growth substrate results in uniaxially aligned 9-AGNRs (GNRs grown along the narrow Au(111) terraces).\citep{Overbeck2019OptimizedSubstratesMeasurement} The Au(788) surface is cleaned in ultrahigh vacuum by two sputtering/annealing cycles: 1~kV Ar$^{+}$ for 10~min followed by annealing at 420~$^\circ$C for 10~min. Next, the precursor monomer DITP is sublimed onto the Au(788) surface from a quartz crucible heated to 70~$^\circ$C, with the substrate held at room temperature. After deposition of about 60\%-70\% of one monolayer DITP, the substrate is heated (0.5~K/s) to 200~$^\circ$C with a 10~min holding time to activate the polymerization reaction, followed by annealing at 400~$^\circ$C (0.5~K/s with a 10~min holding time) to form the GNRs via cyclodehydrogenation.
The average GNR length is between 40 and 45~nm.\citep{DiGiovannantonio2018surfaceGrowthDynamics} 9-AGNRs are transferred from their growth substrate to the silicon-based substrates with predefined graphene electrodes by an electrochemical delamination method using PMMA as described previously.\citep{Overbeck2019OptimizedSubstratesMeasurement, senkovskiy2017making, Overbeck2019UniversalLengthdependent}
As the transfer of the 9-AGNRs onto the substrate exposes the graphene electrodes to water, the samples were heated to 200~$^\circ$C for 120~minutes at 10$^{-6}$~mbar to remove water residues at the graphene/GNR interface and improve the device performance.\citep{braun2021optimized}

\subsection*{Electronic measurements}
All electronic measurements were performed under vacuum conditions ($<$10$^{-6}$~mbar). The devices were measured in a commercially available probe station (Lake Shore Cryogenics, Model CRX-6.5K) at various temperatures (9~K - 350~K). A data acquisition board (ADwin-Gold II, J\"ager Computergesteuerte Messtechnik GmbH) is employed to apply the bias and gate voltages and read the voltage output of the I–V converter (DDPCA-300, FEMTO Messtechnik GmbH).

\subsection*{Theoretical methods}
To model transport through the junctions formed by 9-AGNRs connected to two graphene electrodes, we construct a tight-binding (TB) model of a chain of atoms with one orbital per atom to represent graphene ribbons connected to a 1D chain of atoms as shown in Fig. 11 of the SI. To construct the TB Hamiltonian, all on-site energies are set to zero in the electrodes. The on-site energies for one of the chains representing a ribbon (QD1) are set to zero whereas for the other one (QD2) varies to represent an offset between the energy levels of two ribbons. We assume that the main bias voltage drop happens at the connection point to the electrodes and that the bias potential profile is constant over all sites in the scattering region (see Fig. 11 of the SI). The gate voltage is tuned between -2V and 2V. We employ scattering theory and calculate transmission coefficients for each chain (T$_1$ and T$_2$) representing each ribbon and then use expression 1/T=1/T$_1$+1/T$_2$ to calculate the total transmission (T). To take the effect of addition energy (Coulomb energy) into account, we follow the procedure as in ref \cite{Sadeghi2018TheoryElectronPhonon} and apply constant addition energy U=0.4eV to both QDs. The current then is calculated from the total transmission (T) using the Landauer formula \cite{Sadeghi2018TheoryElectronPhonon} $I(V_b,V_g)=\int T(E,V_b,V_g) (f_L(E,-V_b/2)-f_R(E,V_b/2))$.

\section*{Acknowledgements}
J.Z. acknowledges funding from the EMPAPOSTDOCS-II program which is financed by the European Union Horizon 2020 research and innovation program under the Marie Skłodowska-Curie grant agreement number 754364. O.B. and M.C. acknowledge funding by the EC H2020 FET Open project no. 767187 (QuIET). M.L.P. acknowledges funding from the Swiss National Science Foundation under Spark grant no. 196795 and the Eccellenza Professorial Fellowship no. PCEFP2\textunderscore203663, as well as support by the Swiss State Secretariat for Education, Research and Innovation (SERI) under contract number MB22.00076. M.C. acknowledges funding from the Swiss National Science Foundation under the Sinergia grant no. 189924 (Hydronics). H.S. acknowledges the UKRI for Future Leaders Fellowship number MR/S015329/2. S.S. acknowledges the Leverhulme Trust for Early Career Fellowship no. ECF-2018-375. R.F. acknowledges funding by the Swiss National Science Foundation under grant no. 182015. G.B.B., P.F. and R.F. acknowledge the European Union Horizon 2020 research and innovation program under grant agreement no. 881603 (GrapheneFlagship Core 3) and the Office of Naval Research BRC Program under the grant N00014-18-1-2708. G.B.B., R.D., P.F. and R.F. also greatly appreciate the financial support from the Werner Siemens Foundation (CarboQuant). The authors acknowledge support from the Multiphysics Hub @ Empa for the COMSOL Multiphyics calculations, as well as the Cleanroom Operations Team of the Binnig and Rohrer Nanotechnology Center (BRNC) for their help and support.

\section*{Author contributions}
Conceptualization: M.C., J.Z., O.B., M.L.P, J.O.
Methodology: J.Z., O.B., J.O., S.S., H.S., M.L.P, I.S.
Investigation: J.Z., O.B., M.L.P., J.O., M.S., A.O., H.S., S.S., A.H.S.D.
Resources: R. Fu., G.B.B., R.D., K.M.
Visualization: M.L.P., J.Z., O.B.
Supervision: P.R., R.F., H.S., M.L.P., M.C. 
Writing—original draft: O.B., J.Z., S.S., H.S., M.L.P., M.C.
Writing—review \& editing: J.Z., O.B., M.L.P., M.C., H.S., S.S., A.H.S.D., J.O., I.S., M.S., A.O., G.B.B., R.D., K.M., R.F., P.R.

\section*{Competing interests}
The authors declare that they have no competing interests.

\section*{Data availability}
The datasets generated during and/or analysed during the current study are available from the authors on reasonable request.

\bibliography{references}

\providecommand{\latin}[1]{#1}
\makeatletter
\providecommand{\doi}
  {\begingroup\let\do\@makeother\dospecials
  \catcode`\{=1 \catcode`\}=2 \doi@aux}
\providecommand{\doi@aux}[1]{\endgroup\texttt{#1}}
\makeatother
\providecommand*\mcitethebibliography{\thebibliography}
\csname @ifundefined\endcsname{endmcitethebibliography}
  {\let\endmcitethebibliography\endthebibliography}{}
\begin{mcitethebibliography}{50}
\providecommand*\natexlab[1]{#1}
\providecommand*\mciteSetBstSublistMode[1]{}
\providecommand*\mciteSetBstMaxWidthForm[2]{}
\providecommand*\mciteBstWouldAddEndPuncttrue
  {\def\EndOfBibitem{\unskip.}}
\providecommand*\mciteBstWouldAddEndPunctfalse
  {\let\EndOfBibitem\relax}
\providecommand*\mciteSetBstMidEndSepPunct[3]{}
\providecommand*\mciteSetBstSublistLabelBeginEnd[3]{}
\providecommand*\EndOfBibitem{}
\mciteSetBstSublistMode{f}
\mciteSetBstMaxWidthForm{subitem}{(\alph{mcitesubitemcount})}
\mciteSetBstSublistLabelBeginEnd
  {\mcitemaxwidthsubitemform\space}
  {\relax}
  {\relax}

\bibitem[Gunlycke \latin{et~al.}(2007)Gunlycke, Lawler, and
  White]{gunlycke2007room}
Gunlycke,~D.; Lawler,~H.; White,~C. Room-temperature ballistic transport in
  narrow graphene strips. \emph{Physical Review B} \textbf{2007}, \emph{75},
  085418\relax
\mciteBstWouldAddEndPuncttrue
\mciteSetBstMidEndSepPunct{\mcitedefaultmidpunct}
{\mcitedefaultendpunct}{\mcitedefaultseppunct}\relax
\EndOfBibitem
\bibitem[Stampfer \latin{et~al.}(2009)Stampfer, G{\"u}ttinger, Hellm{\"u}ller,
  Molitor, Ensslin, and Ihn]{stampfer2009energy}
Stampfer,~C.; G{\"u}ttinger,~J.; Hellm{\"u}ller,~S.; Molitor,~F.; Ensslin,~K.;
  Ihn,~T. Energy gaps in etched graphene nanoribbons. \emph{Physical review
  letters} \textbf{2009}, \emph{102}, 056403\relax
\mciteBstWouldAddEndPuncttrue
\mciteSetBstMidEndSepPunct{\mcitedefaultmidpunct}
{\mcitedefaultendpunct}{\mcitedefaultseppunct}\relax
\EndOfBibitem
\bibitem[Wang \latin{et~al.}(2022)Wang, Song, Wang, Guo, Xue, Wang, Wang, Chen,
  Jiang, Chen, Shi, Wu, Song, Zhang, Watanabe, Taniguchi, Song, and
  Xie]{Xiujun2022Minimizing}
Wang,~X.; Song,~S.; Wang,~H.; Guo,~T.; Xue,~Y.; Wang,~R.; Wang,~H.; Chen,~L.;
  Jiang,~C.; Chen,~C.; Shi,~Z.; Wu,~T.; Song,~W.; Zhang,~S.; Watanabe,~K.;
  Taniguchi,~T.; Song,~Z.; Xie,~X. Minimizing the Programming Power of Phase
  Change Memory by Using Graphene Nanoribbon Edge-Contact. \emph{Advanced
  Science} \textbf{2022}, \emph{9}, 2202222\relax
\mciteBstWouldAddEndPuncttrue
\mciteSetBstMidEndSepPunct{\mcitedefaultmidpunct}
{\mcitedefaultendpunct}{\mcitedefaultseppunct}\relax
\EndOfBibitem
\bibitem[Cai \latin{et~al.}(2010)Cai, Ruffieux, Jaafar, Bieri, Braun,
  Blankenburg, Muoth, Seitsonen, Saleh, Feng, M{\"{u}}llen, and
  Fasel]{Cai2010AtomicallyPreciseBottom}
Cai,~J.; Ruffieux,~P.; Jaafar,~R.; Bieri,~M.; Braun,~T.; Blankenburg,~S.;
  Muoth,~M.; Seitsonen,~A.~P.; Saleh,~M.; Feng,~X.; M{\"{u}}llen,~K.; Fasel,~R.
  Atomically Precise Bottom-up Fabrication of Graphene Nanoribbons.
  \emph{Nature} \textbf{2010}, \emph{466}, 470--473\relax
\mciteBstWouldAddEndPuncttrue
\mciteSetBstMidEndSepPunct{\mcitedefaultmidpunct}
{\mcitedefaultendpunct}{\mcitedefaultseppunct}\relax
\EndOfBibitem
\bibitem[Liu \latin{et~al.}({2015})Liu, Li, Tan, Giannakopoulos,
  Sanchez-Sanchez, Beljonne, Ruffieux, Fasel, Feng, and
  Muellen]{Liu2015CoveEdgedLow}
Liu,~J.; Li,~B.-W.; Tan,~Y.-Z.; Giannakopoulos,~A.; Sanchez-Sanchez,~C.;
  Beljonne,~D.; Ruffieux,~P.; Fasel,~R.; Feng,~X.; Muellen,~K. Toward
  Cove-edged Low Band Gap Graphene Nanoribbons. \emph{Journal of the American
  Chemical Society} \textbf{{2015}}, \emph{{137}}, 6097--6103\relax
\mciteBstWouldAddEndPuncttrue
\mciteSetBstMidEndSepPunct{\mcitedefaultmidpunct}
{\mcitedefaultendpunct}{\mcitedefaultseppunct}\relax
\EndOfBibitem
\bibitem[Cai \latin{et~al.}({2014})Cai, Pignedoli, Talirz, Ruffieux, Soede,
  Liang, Meunier, Berger, Li, Feng, Muellen, and
  Fasel]{Cai2014GrapheneNanoribbonHeterojunctions}
Cai,~J.; Pignedoli,~C.~A.; Talirz,~L.; Ruffieux,~P.; Soede,~H.; Liang,~L.;
  Meunier,~V.; Berger,~R.; Li,~R.; Feng,~X.; Muellen,~K.; Fasel,~R. Graphene
  Nanoribbon Heterojunctions. \emph{Nature Nanotechnology} \textbf{{2014}},
  \emph{{9}}, 896--900\relax
\mciteBstWouldAddEndPuncttrue
\mciteSetBstMidEndSepPunct{\mcitedefaultmidpunct}
{\mcitedefaultendpunct}{\mcitedefaultseppunct}\relax
\EndOfBibitem
\bibitem[Ruffieux \latin{et~al.}({2016})Ruffieux, Wang, Yang, Sanchez-Sanchez,
  Liu, Dienel, Talirz, Shinde, Pignedoli, Passerone, Dumslaff, Feng, Muellen,
  and Fasel]{Ruffieux2016surfaceSynthesisGraphene}
Ruffieux,~P.; Wang,~S.; Yang,~B.; Sanchez-Sanchez,~C.; Liu,~J.; Dienel,~T.;
  Talirz,~L.; Shinde,~P.; Pignedoli,~C.~A.; Passerone,~D.; Dumslaff,~T.;
  Feng,~X.; Muellen,~K.; Fasel,~R. On-surface Synthesis of Graphene Nanoribbons
  with Zigzag Edge Topology. \emph{Nature} \textbf{{2016}}, \emph{{531}},
  {489}\relax
\mciteBstWouldAddEndPuncttrue
\mciteSetBstMidEndSepPunct{\mcitedefaultmidpunct}
{\mcitedefaultendpunct}{\mcitedefaultseppunct}\relax
\EndOfBibitem
\bibitem[Groning \latin{et~al.}({2018})Groning, Wang, Yao, Pignedoli, Barin,
  Daniels, Cupo, Meunier, Feng, Narita, Muellen, Ruffieux, and
  Fasel]{Groning2018EngineeringRobustTopological}
Groning,~O.; Wang,~S.; Yao,~X.; Pignedoli,~C.~A.; Barin,~G.~B.; Daniels,~C.;
  Cupo,~A.; Meunier,~V.; Feng,~X.; Narita,~A.; Muellen,~K.; Ruffieux,~P.;
  Fasel,~R. Engineering of Robust Topological Quantum Phases in Graphene
  Nanoribbons. \emph{Nature} \textbf{{2018}}, \emph{{560}}, {209}\relax
\mciteBstWouldAddEndPuncttrue
\mciteSetBstMidEndSepPunct{\mcitedefaultmidpunct}
{\mcitedefaultendpunct}{\mcitedefaultseppunct}\relax
\EndOfBibitem
\bibitem[Rizzo \latin{et~al.}(2018)Rizzo, Veber, Cao, Bronner, Chen, Zhao,
  Rodriguez, Louie, Crommie, and Fischer]{Rizzo2018TopologicalBandEngineering}
Rizzo,~D.~J.; Veber,~G.; Cao,~T.; Bronner,~C.; Chen,~T.; Zhao,~F.;
  Rodriguez,~H.; Louie,~S.~G.; Crommie,~M.~F.; Fischer,~F.~R. Topological Band
  Engineering of Graphene Nanoribbons. \emph{Nature} \textbf{2018}, \emph{560},
  204--208\relax
\mciteBstWouldAddEndPuncttrue
\mciteSetBstMidEndSepPunct{\mcitedefaultmidpunct}
{\mcitedefaultendpunct}{\mcitedefaultseppunct}\relax
\EndOfBibitem
\bibitem[Sun \latin{et~al.}(2020)Sun, Gr{\"{o}}ning, Overbeck, Braun, Perrin,
  {Borin Barin}, {El Abbassi}, Eimre, Ditler, Daniels, Meunier, Pignedoli,
  Calame, Fasel, and Ruffieux]{Sun2020MassiveDiracFermion}
Sun,~Q.; Gr{\"{o}}ning,~O.; Overbeck,~J.; Braun,~O.; Perrin,~M.~L.; {Borin
  Barin},~G.; {El Abbassi},~M.; Eimre,~K.; Ditler,~E.; Daniels,~C.;
  Meunier,~V.; Pignedoli,~C.~A.; Calame,~M.; Fasel,~R.; Ruffieux,~P. Massive
  Dirac Fermion Behavior in a Low Bandgap Graphene Nanoribbon near a
  Topological Phase Boundary. \emph{Advanced Materials} \textbf{2020},
  \emph{32}, 1906054\relax
\mciteBstWouldAddEndPuncttrue
\mciteSetBstMidEndSepPunct{\mcitedefaultmidpunct}
{\mcitedefaultendpunct}{\mcitedefaultseppunct}\relax
\EndOfBibitem
\bibitem[Yamaguchi \latin{et~al.}(2020)Yamaguchi, Hayashi, Jippo, Shiotari,
  Ohtomo, Sakakura, Hieda, Aratani, Ohfuchi, Sugimoto, Yamada, and
  Sato]{Yamaguchi2020Smallbandgapatomically}
Yamaguchi,~J.; Hayashi,~H.; Jippo,~H.; Shiotari,~A.; Ohtomo,~M.; Sakakura,~M.;
  Hieda,~N.; Aratani,~N.; Ohfuchi,~M.; Sugimoto,~Y.; Yamada,~H.; Sato,~S. Small
  Bandgap in Atomically Precise 17-atom-wide Armchair-edged Graphene
  Nanoribbons. \emph{Communications Materials} \textbf{2020}, \emph{1},
  36\relax
\mciteBstWouldAddEndPuncttrue
\mciteSetBstMidEndSepPunct{\mcitedefaultmidpunct}
{\mcitedefaultendpunct}{\mcitedefaultseppunct}\relax
\EndOfBibitem
\bibitem[Chen \latin{et~al.}(2020)Chen, Narita, and
  MÃ¼llen]{Chen2020GrapheneNanoribbonssurface}
Chen,~Z.; Narita,~A.; MÃ¼llen,~K. Graphene Nanoribbons: On-surface Synthesis
  and Integration into Electronic Devices. \emph{Advanced Materials}
  \textbf{2020}, \emph{32}, 2001893\relax
\mciteBstWouldAddEndPuncttrue
\mciteSetBstMidEndSepPunct{\mcitedefaultmidpunct}
{\mcitedefaultendpunct}{\mcitedefaultseppunct}\relax
\EndOfBibitem
\bibitem[Li \latin{et~al.}(2020)Li, Zee, Lin, Basile, Muni, Flores,
  MunÃ¡rriz, Kaner, Alexandrova, Houk, Tolbert, and
  Rubin]{Li2020FjordEdgeGraphene}
Li,~Y.~L.; Zee,~C.-T.; Lin,~J.~B.; Basile,~V.~M.; Muni,~M.; Flores,~M.~D.;
  MunÃ¡rriz,~J.; Kaner,~R.~B.; Alexandrova,~A.~N.; Houk,~K.~N.;
  Tolbert,~S.~H.; Rubin,~Y. Fjord-edge Graphene Nanoribbons with Site-specific
  Nitrogen Substitution. \emph{Journal of the American Chemical Society}
  \textbf{2020}, \emph{142}, 18093--18102\relax
\mciteBstWouldAddEndPuncttrue
\mciteSetBstMidEndSepPunct{\mcitedefaultmidpunct}
{\mcitedefaultendpunct}{\mcitedefaultseppunct}\relax
\EndOfBibitem
\bibitem[Pawlak \latin{et~al.}(2020)Pawlak, Liu, Ninova, Dâ€™Astolfo,
  Drechsel, Sangtarash, HÃ¤ner, Decurtins, Sadeghi, Lambert, Aschauer, Liu,
  and Meyer]{Pawlak2020BottomSynthesisNitrogen}
Pawlak,~R.; Liu,~X.; Ninova,~S.; Dâ€™Astolfo,~P.; Drechsel,~C.;
  Sangtarash,~S.; HÃ¤ner,~R.; Decurtins,~S.; Sadeghi,~H.; Lambert,~C.~J.;
  Aschauer,~U.; Liu,~S.-X.; Meyer,~E. Bottom-up Synthesis of Nitrogen-doped
  Porous Graphene Nanoribbons. \emph{Journal of the American Chemical Society}
  \textbf{2020}, \emph{142}, 12568--12573\relax
\mciteBstWouldAddEndPuncttrue
\mciteSetBstMidEndSepPunct{\mcitedefaultmidpunct}
{\mcitedefaultendpunct}{\mcitedefaultseppunct}\relax
\EndOfBibitem
\bibitem[Sun \latin{et~al.}(2020)Sun, Yao, GrÃ¶ning, Eimre, Pignedoli,
  MÃ¼llen, Narita, Fasel, and Ruffieux]{Sun2020CoupledSpinStates}
Sun,~Q.; Yao,~X.; GrÃ¶ning,~O.; Eimre,~K.; Pignedoli,~C.~A.; MÃ¼llen,~K.;
  Narita,~A.; Fasel,~R.; Ruffieux,~P. Coupled Spin States in Armchair Graphene
  Nanoribbons with Asymmetric Zigzag Edge Extensions. \emph{Nano Letters}
  \textbf{2020}, \emph{20}, 6429--6436\relax
\mciteBstWouldAddEndPuncttrue
\mciteSetBstMidEndSepPunct{\mcitedefaultmidpunct}
{\mcitedefaultendpunct}{\mcitedefaultseppunct}\relax
\EndOfBibitem
\bibitem[Cirera \latin{et~al.}(2020)Cirera, Sánchez-Grande, de~la Torre,
  Santos, Edalatmanesh, Rodríguez-Sánchez, Lauwaet, Mallada, Zbořil,
  Miranda, Gröning, Jelínek, Martín, and
  Ecija]{Cirera2020TailoringTopologicalOrder}
Cirera,~B.; Sánchez-Grande,~A.; de~la Torre,~B.; Santos,~J.; Edalatmanesh,~S.;
  Rodríguez-Sánchez,~E.; Lauwaet,~K.; Mallada,~B.; Zbořil,~R.; Miranda,~R.;
  Gröning,~O.; Jelínek,~P.; Martín,~N.; Ecija,~D. Tailoring Topological
  Order and $\pi$-conjugation to Engineer Quasi-metallic Polymers. \emph{Nature
  Nanotechnology} \textbf{2020}, \emph{15}, 437--443\relax
\mciteBstWouldAddEndPuncttrue
\mciteSetBstMidEndSepPunct{\mcitedefaultmidpunct}
{\mcitedefaultendpunct}{\mcitedefaultseppunct}\relax
\EndOfBibitem
\bibitem[Yang \latin{et~al.}(2022)Yang, Ma, Zheng, Osella, Droste, Komber, Liu,
  Böckmann, Beljonne, Hansen, Bonn, Wang, Liu, and Feng]{Yang2022Solution}
Yang,~L.; Ma,~J.; Zheng,~W.; Osella,~S.; Droste,~J.; Komber,~H.; Liu,~K.;
  Böckmann,~S.; Beljonne,~D.; Hansen,~M.~R.; Bonn,~M.; Wang,~H.~I.; Liu,~J.;
  Feng,~X. Solution Synthesis and Characterization of a Long and Curved
  Graphene Nanoribbon with Hybrid Cove–Armchair–Gulf Edge Structures.
  \emph{Advanced Science} \textbf{2022}, \emph{9}, 2200708\relax
\mciteBstWouldAddEndPuncttrue
\mciteSetBstMidEndSepPunct{\mcitedefaultmidpunct}
{\mcitedefaultendpunct}{\mcitedefaultseppunct}\relax
\EndOfBibitem
\bibitem[Son \latin{et~al.}(2006)Son, Cohen, and
  Louie]{Son2006EnergyGapsGraphene}
Son,~Y.-W.; Cohen,~M.~L.; Louie,~S.~G. Energy Gaps in Graphene Nanoribbons.
  \emph{Physical Review Letters} \textbf{2006}, \emph{97}, 216803\relax
\mciteBstWouldAddEndPuncttrue
\mciteSetBstMidEndSepPunct{\mcitedefaultmidpunct}
{\mcitedefaultendpunct}{\mcitedefaultseppunct}\relax
\EndOfBibitem
\bibitem[Kimouche \latin{et~al.}(2015)Kimouche, Ervasti, Drost, Halonen, Harju,
  Joensuu, Sainio, and Liljeroth]{Kimouche2015Ultranarrowmetallic}
Kimouche,~A.; Ervasti,~M.~M.; Drost,~R.; Halonen,~S.; Harju,~A.;
  Joensuu,~P.~M.; Sainio,~J.; Liljeroth,~P. Ultra-narrow Metallic Armchair
  Graphene Nanoribbons. \emph{Nature Communications} \textbf{2015}, \emph{6},
  10177\relax
\mciteBstWouldAddEndPuncttrue
\mciteSetBstMidEndSepPunct{\mcitedefaultmidpunct}
{\mcitedefaultendpunct}{\mcitedefaultseppunct}\relax
\EndOfBibitem
\bibitem[Wang \latin{et~al.}(2016)Wang, Talirz, Pignedoli, Feng, M{\"u}llen,
  Fasel, and Ruffieux]{wang2016giant}
Wang,~S.; Talirz,~L.; Pignedoli,~C.~A.; Feng,~X.; M{\"u}llen,~K.; Fasel,~R.;
  Ruffieux,~P. Giant edge state splitting at atomically precise graphene zigzag
  edges. \emph{Nature communications} \textbf{2016}, \emph{7}, 1--6\relax
\mciteBstWouldAddEndPuncttrue
\mciteSetBstMidEndSepPunct{\mcitedefaultmidpunct}
{\mcitedefaultendpunct}{\mcitedefaultseppunct}\relax
\EndOfBibitem
\bibitem[Cao \latin{et~al.}(2017)Cao, Zhao, and
  Louie]{Cao2017TopologicalPhasesGraphene}
Cao,~T.; Zhao,~F.; Louie,~S.~G. Topological Phases in Graphene Nanoribbons:
  Junction States, Spin Centers, and Quantum Spin Chains. \emph{Physical Review
  Letters} \textbf{2017}, \emph{119}, 076401\relax
\mciteBstWouldAddEndPuncttrue
\mciteSetBstMidEndSepPunct{\mcitedefaultmidpunct}
{\mcitedefaultendpunct}{\mcitedefaultseppunct}\relax
\EndOfBibitem
\bibitem[Llinas \latin{et~al.}(2017)Llinas, Fairbrother, {Borin Barin}, Shi,
  Lee, Wu, {Yong Choi}, Braganza, Lear, Kau, Choi, Chen, Pedramrazi, Dumslaff,
  Narita, Feng, M{\"{u}}llen, Fischer, Zettl, Ruffieux, Yablonovitch, Crommie,
  Fasel, and Bokor]{Llinas2017ShortchannelField}
Llinas,~J.~P.; Fairbrother,~A.; {Borin Barin},~G.; Shi,~W.; Lee,~K.; Wu,~S.;
  {Yong Choi},~B.; Braganza,~R.; Lear,~J.; Kau,~N.; Choi,~W.; Chen,~C.;
  Pedramrazi,~Z.; Dumslaff,~T.; Narita,~A.; Feng,~X.; M{\"{u}}llen,~K.;
  Fischer,~F.; Zettl,~A.; Ruffieux,~P. \latin{et~al.}  Short-channel
  Field-effect Transistors with 9-atom and 13-atom Wide Graphene Nanoribbons.
  \emph{Nature Communications} \textbf{2017}, \emph{8}, 633\relax
\mciteBstWouldAddEndPuncttrue
\mciteSetBstMidEndSepPunct{\mcitedefaultmidpunct}
{\mcitedefaultendpunct}{\mcitedefaultseppunct}\relax
\EndOfBibitem
\bibitem[Martini \latin{et~al.}(2019)Martini, Chen, Mishra, Barin, Fantuzzi,
  Ruffieux, Fasel, Feng, Narita, Coletti, M{\"{u}}llen, and
  Candini]{Martini2019StructuredependentElectrical}
Martini,~L.; Chen,~Z.; Mishra,~N.; Barin,~G.~B.; Fantuzzi,~P.; Ruffieux,~P.;
  Fasel,~R.; Feng,~X.; Narita,~A.; Coletti,~C.; M{\"{u}}llen,~K.; Candini,~A.
  Structure-dependent Electrical Properties of Graphene Nanoribbon Devices with
  Graphene Electrodes. \emph{Carbon} \textbf{2019}, \emph{146}, 36--43\relax
\mciteBstWouldAddEndPuncttrue
\mciteSetBstMidEndSepPunct{\mcitedefaultmidpunct}
{\mcitedefaultendpunct}{\mcitedefaultseppunct}\relax
\EndOfBibitem
\bibitem[Passi \latin{et~al.}(2018)Passi, Gahoi, Senkovskiy, Haberer, Fischer,
  Gr{\"{u}}neis, and Lemme]{Passi2018FieldeffectTransistors}
Passi,~V.; Gahoi,~A.; Senkovskiy,~B.~V.; Haberer,~D.; Fischer,~F.~R.;
  Gr{\"{u}}neis,~A.; Lemme,~M.~C. Field-effect Transistors Based on Networks of
  Highly Aligned, Chemically Synthesized N = 7 Armchair Graphene Nanoribbons.
  \emph{{ACS Applied Materials and Interfaces}} \textbf{2018}, \emph{10},
  9900--9903\relax
\mciteBstWouldAddEndPuncttrue
\mciteSetBstMidEndSepPunct{\mcitedefaultmidpunct}
{\mcitedefaultendpunct}{\mcitedefaultseppunct}\relax
\EndOfBibitem
\bibitem[{El Abbassi} \latin{et~al.}(2020){El Abbassi}, Perrin, Barin,
  Sangtarash, Overbeck, Braun, Lambert, Sun, Prechtl, Narita, M{\"{u}}llen,
  Ruffieux, Sadeghi, Fasel, and Calame]{ElAbbassi2020ControlledQuantumDot}
{El Abbassi},~M.; Perrin,~M.~L.; Barin,~G.~B.; Sangtarash,~S.; Overbeck,~J.;
  Braun,~O.; Lambert,~C.~J.; Sun,~Q.; Prechtl,~T.; Narita,~A.;
  M{\"{u}}llen,~K.; Ruffieux,~P.; Sadeghi,~H.; Fasel,~R.; Calame,~M. Controlled
  Quantum Dot Formation in Atomically Engineered Graphene Nanoribbon
  Field-effect Transistors. \emph{{{ACS Nano}}} \textbf{2020}, \emph{14},
  5754--5762\relax
\mciteBstWouldAddEndPuncttrue
\mciteSetBstMidEndSepPunct{\mcitedefaultmidpunct}
{\mcitedefaultendpunct}{\mcitedefaultseppunct}\relax
\EndOfBibitem
\bibitem[Richter \latin{et~al.}(2020)Richter, Chen, Tries, Prechtl, Narita,
  M{\"{u}}llen, Asadi, Bonn, and
  Kl{\"{a}}ui]{Richter2020ChargeTransportMechanism}
Richter,~N.; Chen,~Z.; Tries,~A.; Prechtl,~T.; Narita,~A.; M{\"{u}}llen,~K.;
  Asadi,~K.; Bonn,~M.; Kl{\"{a}}ui,~M. Charge Transport Mechanism in Networks
  of Armchair Graphene Nanoribbons. \emph{Scientific Reports} \textbf{2020},
  \emph{10}, 1--8\relax
\mciteBstWouldAddEndPuncttrue
\mciteSetBstMidEndSepPunct{\mcitedefaultmidpunct}
{\mcitedefaultendpunct}{\mcitedefaultseppunct}\relax
\EndOfBibitem
\bibitem[Braun \latin{et~al.}(2021)Braun, Overbeck, Abbassi, K{\"a}ser, Furrer,
  Olziersky, Flasby, Barin, Darawish, M{\"u}llen, \latin{et~al.}
  others]{braun2021optimized}
Braun,~O.; Overbeck,~J.; Abbassi,~M.~E.; K{\"a}ser,~S.; Furrer,~R.;
  Olziersky,~A.; Flasby,~A.; Barin,~G.~B.; Darawish,~R.; M{\"u}llen,~K.,
  \latin{et~al.}  Optimized Graphene Electrodes for contacting Graphene
  Nanoribbons. \emph{Carbon} \textbf{2021}, \emph{184}, 331--339\relax
\mciteBstWouldAddEndPuncttrue
\mciteSetBstMidEndSepPunct{\mcitedefaultmidpunct}
{\mcitedefaultendpunct}{\mcitedefaultseppunct}\relax
\EndOfBibitem
\bibitem[Senkovskiy \latin{et~al.}(2021)Senkovskiy, Nenashev, Alavi, Falke,
  Hell, Bampoulis, Rybkovskiy, Usachov, Fedorov, Chernov, \latin{et~al.}
  others]{senkovskiy2021tunneling}
Senkovskiy,~B.~V.; Nenashev,~A.~V.; Alavi,~S.~K.; Falke,~Y.; Hell,~M.;
  Bampoulis,~P.; Rybkovskiy,~D.~V.; Usachov,~D.~Y.; Fedorov,~A.~V.;
  Chernov,~A.~I., \latin{et~al.}  Tunneling current modulation in atomically
  precise graphene nanoribbon heterojunctions. \emph{Nature communications}
  \textbf{2021}, \emph{12}, 1--11\relax
\mciteBstWouldAddEndPuncttrue
\mciteSetBstMidEndSepPunct{\mcitedefaultmidpunct}
{\mcitedefaultendpunct}{\mcitedefaultseppunct}\relax
\EndOfBibitem
\bibitem[Saraswat \latin{et~al.}(2021)Saraswat, Jacobberger, and
  Arnold]{saraswat2021materials}
Saraswat,~V.; Jacobberger,~R.~M.; Arnold,~M.~S. Materials Science Challenges to
  Graphene Nanoribbon Electronics. \emph{ACS nano} \textbf{2021}, \emph{15},
  3674--3708\relax
\mciteBstWouldAddEndPuncttrue
\mciteSetBstMidEndSepPunct{\mcitedefaultmidpunct}
{\mcitedefaultendpunct}{\mcitedefaultseppunct}\relax
\EndOfBibitem
\bibitem[Wu \latin{et~al.}(2022)Wu, Tian, Shen, Hou, Ren, Gou, Sun, Yang, and
  Ren]{wu2022vertical}
Wu,~F.; Tian,~H.; Shen,~Y.; Hou,~Z.; Ren,~J.; Gou,~G.; Sun,~Y.; Yang,~Y.;
  Ren,~T.-L. Vertical MoS2 transistors with sub-1-nm gate lengths.
  \emph{Nature} \textbf{2022}, \emph{603}, 259--264\relax
\mciteBstWouldAddEndPuncttrue
\mciteSetBstMidEndSepPunct{\mcitedefaultmidpunct}
{\mcitedefaultendpunct}{\mcitedefaultseppunct}\relax
\EndOfBibitem
\bibitem[Jiang \latin{et~al.}(2020)Jiang, Doan, Sun, Kim, Yu, Joo, Park, Yang,
  Duong, and Lee]{Jiang2020Ultrashort}
Jiang,~J.; Doan,~M.-H.; Sun,~L.; Kim,~H.; Yu,~H.; Joo,~M.-K.; Park,~S.~H.;
  Yang,~H.; Duong,~D.~L.; Lee,~Y.~H. Ultrashort Vertical-Channel van der Waals
  Semiconductor Transistors. \emph{Advanced Science} \textbf{2020}, \emph{7},
  1902964\relax
\mciteBstWouldAddEndPuncttrue
\mciteSetBstMidEndSepPunct{\mcitedefaultmidpunct}
{\mcitedefaultendpunct}{\mcitedefaultseppunct}\relax
\EndOfBibitem
\bibitem[El~Abbassi \latin{et~al.}(2019)El~Abbassi, Sangtarash, Liu, Perrin,
  Braun, Lambert, van~der Zant, Yitzchaik, Decurtins, Liu, Sadeghi, and
  Calame]{ElAbbassi2019RobustGraphenebased}
El~Abbassi,~M.; Sangtarash,~S.; Liu,~X.; Perrin,~M.~L.; Braun,~O.; Lambert,~C.;
  van~der Zant,~H. S.~J.; Yitzchaik,~S.; Decurtins,~S.; Liu,~S.-X.;
  Sadeghi,~H.; Calame,~M. Robust Graphene-based Molecular Devices. \emph{Nature
  Nanotechnology} \textbf{2019}, \emph{14}, 957--961\relax
\mciteBstWouldAddEndPuncttrue
\mciteSetBstMidEndSepPunct{\mcitedefaultmidpunct}
{\mcitedefaultendpunct}{\mcitedefaultseppunct}\relax
\EndOfBibitem
\bibitem[Radisavljevic \latin{et~al.}(2011)Radisavljevic, Radenovic, Brivio,
  Giacometti, and Kis]{radisavljevic2011single}
Radisavljevic,~B.; Radenovic,~A.; Brivio,~J.; Giacometti,~V.; Kis,~A.
  Single-layer MoS 2 transistors. \emph{Nature nanotechnology} \textbf{2011},
  \emph{6}, 147--150\relax
\mciteBstWouldAddEndPuncttrue
\mciteSetBstMidEndSepPunct{\mcitedefaultmidpunct}
{\mcitedefaultendpunct}{\mcitedefaultseppunct}\relax
\EndOfBibitem
\bibitem[Marega \latin{et~al.}(2020)Marega, Zhao, Avsar, Wang, Tripathi,
  Radenovic, and Kis]{marega2020logic}
Marega,~G.~M.; Zhao,~Y.; Avsar,~A.; Wang,~Z.; Tripathi,~M.; Radenovic,~A.;
  Kis,~A. Logic-in-memory based on an atomically thin semiconductor.
  \emph{Nature} \textbf{2020}, \emph{587}, 72--77\relax
\mciteBstWouldAddEndPuncttrue
\mciteSetBstMidEndSepPunct{\mcitedefaultmidpunct}
{\mcitedefaultendpunct}{\mcitedefaultseppunct}\relax
\EndOfBibitem
\bibitem[Liu \latin{et~al.}(2021)Liu, Kong, Li, He, Ren, Tao, Yang, Lin, Zhao,
  Li, \latin{et~al.} others]{liu2021transferred}
Liu,~L.; Kong,~L.; Li,~Q.; He,~C.; Ren,~L.; Tao,~Q.; Yang,~X.; Lin,~J.;
  Zhao,~B.; Li,~Z., \latin{et~al.}  Transferred van der Waals metal electrodes
  for sub-1-nm MoS2 vertical transistors. \emph{Nature Electronics}
  \textbf{2021}, \emph{4}, 342--347\relax
\mciteBstWouldAddEndPuncttrue
\mciteSetBstMidEndSepPunct{\mcitedefaultmidpunct}
{\mcitedefaultendpunct}{\mcitedefaultseppunct}\relax
\EndOfBibitem
\bibitem[Wu \latin{et~al.}(2007)Wu, Lin, Ye, and Wilk]{wu2007current}
Wu,~Y.; Lin,~H.; Ye,~P.; Wilk,~G. Current transport and maximum dielectric
  strength of atomic-layer-deposited ultrathin Al 2 O 3 on GaAs. \emph{Applied
  physics letters} \textbf{2007}, \emph{90}, 072105\relax
\mciteBstWouldAddEndPuncttrue
\mciteSetBstMidEndSepPunct{\mcitedefaultmidpunct}
{\mcitedefaultendpunct}{\mcitedefaultseppunct}\relax
\EndOfBibitem
\bibitem[Di~Giovannantonio \latin{et~al.}(2018)Di~Giovannantonio, Deniz, Urgel,
  Widmer, Dienel, Stolz, S{\'a}nchez-S{\'a}nchez, Muntwiler, Dumslaff, Berger,
  Narita, Feng, M{\"u}llen, Ruffieux, and
  Fasel]{DiGiovannantonio2018surfaceGrowthDynamics}
Di~Giovannantonio,~M.; Deniz,~O.; Urgel,~J.~I.; Widmer,~R.; Dienel,~T.;
  Stolz,~S.; S{\'a}nchez-S{\'a}nchez,~C.; Muntwiler,~M.; Dumslaff,~T.;
  Berger,~R.; Narita,~A.; Feng,~X.; M{\"u}llen,~K.; Ruffieux,~P.; Fasel,~R.
  On-surface Growth Dynamics of Graphene Nanoribbons: The Role of Halogen
  Functionalization. \emph{{ACS Nano}} \textbf{2018}, \emph{12}, 74--81\relax
\mciteBstWouldAddEndPuncttrue
\mciteSetBstMidEndSepPunct{\mcitedefaultmidpunct}
{\mcitedefaultendpunct}{\mcitedefaultseppunct}\relax
\EndOfBibitem
\bibitem[Overbeck \latin{et~al.}(2019)Overbeck, {Borin Barin}, Daniels, Perrin,
  Liang, Braun, Darawish, Burkhardt, Dumslaff, Wang, Narita, M{\"{u}}llen,
  Meunier, Fasel, Calame, and
  Ruffieux]{Overbeck2019OptimizedSubstratesMeasurement}
Overbeck,~J.; {Borin Barin},~G.; Daniels,~C.; Perrin,~M.~L.; Liang,~L.;
  Braun,~O.; Darawish,~R.; Burkhardt,~B.; Dumslaff,~T.; Wang,~X.~Y.;
  Narita,~A.; M{\"{u}}llen,~K.; Meunier,~V.; Fasel,~R.; Calame,~M.;
  Ruffieux,~P. Optimized Substrates and Measurement Approaches for Raman
  Spectroscopy of Graphene Nanoribbons. \emph{Physica Status Solidi (b) Basic
  Research} \textbf{2019}, \emph{256}, 1900343\relax
\mciteBstWouldAddEndPuncttrue
\mciteSetBstMidEndSepPunct{\mcitedefaultmidpunct}
{\mcitedefaultendpunct}{\mcitedefaultseppunct}\relax
\EndOfBibitem
\bibitem[Senkovskiy \latin{et~al.}(2017)Senkovskiy, Pfeiffer, Alavi, Bliesener,
  Zhu, Michel, Fedorov, German, Hertel, Haberer, Petaccia, Fischer, Meerholz,
  van Loosdrecht, Lindfors, and Grüneis]{senkovskiy2017making}
Senkovskiy,~B.; Pfeiffer,~M.; Alavi,~S.; Bliesener,~A.; Zhu,~J.; Michel,~S.;
  Fedorov,~A.; German,~R.; Hertel,~D.; Haberer,~D.; Petaccia,~L.; Fischer,~F.;
  Meerholz,~K.; van Loosdrecht,~P.; Lindfors,~K.; Grüneis,~A. Making graphene
  nanoribbons photoluminescent. \emph{Nano letters} \textbf{2017}, \emph{17},
  4029--4037\relax
\mciteBstWouldAddEndPuncttrue
\mciteSetBstMidEndSepPunct{\mcitedefaultmidpunct}
{\mcitedefaultendpunct}{\mcitedefaultseppunct}\relax
\EndOfBibitem
\bibitem[Overbeck \latin{et~al.}(2019)Overbeck, Barin, Daniels, Perrin, Braun,
  Sun, Darawish, {De Luca}, Wang, Dumslaff, Narita, M{\"{u}}llen, Ruffieux,
  Meunier, Fasel, and Calame]{Overbeck2019UniversalLengthdependent}
Overbeck,~J.; Barin,~G.~B.; Daniels,~C.; Perrin,~M.~L.; Braun,~O.; Sun,~Q.;
  Darawish,~R.; {De Luca},~M.; Wang,~X.~Y.; Dumslaff,~T.; Narita,~A.;
  M{\"{u}}llen,~K.; Ruffieux,~P.; Meunier,~V.; Fasel,~R.; Calame,~M. A
  Universal Length-dependent Vibrational Mode in Graphene Nanoribbons.
  \emph{{ACS Nano}} \textbf{2019}, \emph{13}, 13083--13091\relax
\mciteBstWouldAddEndPuncttrue
\mciteSetBstMidEndSepPunct{\mcitedefaultmidpunct}
{\mcitedefaultendpunct}{\mcitedefaultseppunct}\relax
\EndOfBibitem
\bibitem[Kouwenhoven \latin{et~al.}(1997)Kouwenhoven, Marcus, McEuen, Tarucha,
  Westervelt, and Wingreen]{Kouwenhoven1997}
Kouwenhoven,~L.~P.; Marcus,~C.~M.; McEuen,~P.~L.; Tarucha,~S.;
  Westervelt,~R.~M.; Wingreen,~N.~S. In \emph{Mesoscopic Electron Transport};
  Sohn,~L.~L., Kouwenhoven,~L.~P., Sch{\"o}n,~G., Eds.; Springer Netherlands:
  Dordrecht, 1997; pp 105--214\relax
\mciteBstWouldAddEndPuncttrue
\mciteSetBstMidEndSepPunct{\mcitedefaultmidpunct}
{\mcitedefaultendpunct}{\mcitedefaultseppunct}\relax
\EndOfBibitem
\bibitem[Merino-Díez \latin{et~al.}(2017)Merino-Díez, Garcia-Lekue,
  Carbonell-Sanromà, Li, Corso, Colazzo, Sedona, Sánchez-Portal, Pascual, and
  de~Oteyza]{MerinoDiez2017WidthdependentBand}
Merino-Díez,~N.; Garcia-Lekue,~A.; Carbonell-Sanromà,~E.; Li,~J.; Corso,~M.;
  Colazzo,~L.; Sedona,~F.; Sánchez-Portal,~D.; Pascual,~J.~I.;
  de~Oteyza,~D.~G. Width-dependent Band Gap in Armchair Graphene Nanoribbons
  Reveals Fermi Level Pinning on Au(111). \emph{ACS Nano} \textbf{2017},
  \emph{11}, 11661--11668, PMID: 29049879\relax
\mciteBstWouldAddEndPuncttrue
\mciteSetBstMidEndSepPunct{\mcitedefaultmidpunct}
{\mcitedefaultendpunct}{\mcitedefaultseppunct}\relax
\EndOfBibitem
\bibitem[Pizzochero \latin{et~al.}(2021)Pizzochero,
  {\v{C}}er{\c{n}}evi{\v{c}}s, Barin, Wang, Ruffieux, Fasel, and
  Yazyev]{Pizzochero2021Quantumelectronictransport}
Pizzochero,~M.; {\v{C}}er{\c{n}}evi{\v{c}}s,~K.; Barin,~G.~B.; Wang,~S.;
  Ruffieux,~P.; Fasel,~R.; Yazyev,~O.~V. Quantum electronic transport across
  `bite' defects in graphene nanoribbons. \emph{2D Materials} \textbf{2021},
  \emph{8}, 035025\relax
\mciteBstWouldAddEndPuncttrue
\mciteSetBstMidEndSepPunct{\mcitedefaultmidpunct}
{\mcitedefaultendpunct}{\mcitedefaultseppunct}\relax
\EndOfBibitem
\bibitem[Ionica \latin{et~al.}()Ionica, Mont{\`e}s, Zimmermann, Saminadayar,
  and Bouchiat]{ionicainfluence}
Ionica,~I.; Mont{\`e}s,~L.; Zimmermann,~J.; Saminadayar,~L.; Bouchiat,~V.
  Influence of the Geometry on the Coulomb Blockade in Silicon Nanostructures.
  \relax
\mciteBstWouldAddEndPunctfalse
\mciteSetBstMidEndSepPunct{\mcitedefaultmidpunct}
{}{\mcitedefaultseppunct}\relax
\EndOfBibitem
\bibitem[Sadeghi(2018)]{Sadeghi2018TheoryElectronPhonon}
Sadeghi,~H. Theory of Electron, Phonon and Spin Transport in Nanoscale Quantum
  Devices. \emph{Nanotechnology} \textbf{2018}, \emph{29}, 373001\relax
\mciteBstWouldAddEndPuncttrue
\mciteSetBstMidEndSepPunct{\mcitedefaultmidpunct}
{\mcitedefaultendpunct}{\mcitedefaultseppunct}\relax
\EndOfBibitem
\bibitem[Talirz \latin{et~al.}({2017})Talirz, Sode, Dumslaff, Wang,
  Sanchez-Valencia, Liu, Shinde, Pignedoli, Liang, Meunier, Plumb, Shi, Feng,
  Narita, Muellen, Fasel, and
  Ruffieux]{Talirz2017surfaceSynthesisCharacterization}
Talirz,~L.; Sode,~H.; Dumslaff,~T.; Wang,~S.; Sanchez-Valencia,~J.~R.; Liu,~J.;
  Shinde,~P.; Pignedoli,~C.~A.; Liang,~L.; Meunier,~V.; Plumb,~N.~C.; Shi,~M.;
  Feng,~X.; Narita,~A.; Muellen,~K.; Fasel,~R.; Ruffieux,~P. On-surface
  Synthesis and Characterization of 9-atom Wide Armchair Graphene Nanoribbons.
  \emph{{ACS Nano}} \textbf{{2017}}, \emph{{11}}, 1380--1388\relax
\mciteBstWouldAddEndPuncttrue
\mciteSetBstMidEndSepPunct{\mcitedefaultmidpunct}
{\mcitedefaultendpunct}{\mcitedefaultseppunct}\relax
\EndOfBibitem
\bibitem[Zhang \latin{et~al.}(2022)Zhang, Calame, and
  Perrin]{Zhang2022Contactingatomicallyprecise}
Zhang,~J.; Calame,~M.; Perrin,~M.~L. Contacting atomically precise graphene
  nanoribbons for next-generation quantum electronics. \emph{Matter}
  \textbf{2022}, \emph{5}, 2497--2499\relax
\mciteBstWouldAddEndPuncttrue
\mciteSetBstMidEndSepPunct{\mcitedefaultmidpunct}
{\mcitedefaultendpunct}{\mcitedefaultseppunct}\relax
\EndOfBibitem
\bibitem[Desai \latin{et~al.}(2016)Desai, Madhvapathy, Sachid, Llinas, Wang,
  Ahn, Pitner, Kim, Bokor, Hu, Wong, and Javey]{desai2016mos2}
Desai,~S.~B.; Madhvapathy,~S.~R.; Sachid,~A.~B.; Llinas,~J.~P.; Wang,~Q.;
  Ahn,~G.~H.; Pitner,~G.; Kim,~M.~J.; Bokor,~J.; Hu,~C.; Wong,~H.-S.~P.;
  Javey,~A. MoS2 transistors with 1-nanometer gate lengths. \emph{Science}
  \textbf{2016}, \emph{354}, 99--102\relax
\mciteBstWouldAddEndPuncttrue
\mciteSetBstMidEndSepPunct{\mcitedefaultmidpunct}
{\mcitedefaultendpunct}{\mcitedefaultseppunct}\relax
\EndOfBibitem
\bibitem[Wang \latin{et~al.}(2021)Wang, Wang, Ma, Chen, Jiang, Chen, Xie, Li,
  and Wang]{wang2021graphene}
Wang,~H.; Wang,~H.~S.; Ma,~C.; Chen,~L.; Jiang,~C.; Chen,~C.; Xie,~X.;
  Li,~A.-P.; Wang,~X. Graphene nanoribbons for quantum electronics.
  \emph{Nature Reviews Physics} \textbf{2021}, \emph{3}, 791--802\relax
\mciteBstWouldAddEndPuncttrue
\mciteSetBstMidEndSepPunct{\mcitedefaultmidpunct}
{\mcitedefaultendpunct}{\mcitedefaultseppunct}\relax
\EndOfBibitem
\end{mcitethebibliography}

\end{document}